\documentclass[aps,pra,twocolumn,showpacs,balancelastpage,superscriptaddress,groupedaddress]{revtex4}  
\usepackage[T1]{fontenc}
\usepackage[latin1]{inputenc}
\usepackage[english]{babel}
\usepackage[full]{complexity}
\usepackage{multirow}

\usepackage{graphicx,subfigure}  
\usepackage{dcolumn}   
\usepackage{bm}        
\usepackage{amssymb}   
\input{Qcircuit.tex}
\renewcommand{\Qcircuit}[1][0em]{\xymatrix @*=<#1>}
\newcommand{\floor}[1]{\left\lfloor #1 \right\rfloor}
\newcommand{\ceil}[1]{\left\lceil #1 \right\rceil}

\begin{document}

\title{\large Quantum Circuits for GCD Computation with $O(n \log n)$ Depth and $O(n)$ Ancillae}
\author{Mehdi Saeedi}
\email[Address correspondence to: ]{msaeedi@usc.edu}
\affiliation{Department of Electrical Engineering, University of Southern California, Los Angeles, CA 90089-2562}

\author{Igor L. Markov}
\affiliation{Department of EECS, University of Michigan, Ann Arbor, MI 48109-2121}

\begin{abstract}
GCD computations and variants of the Euclidean algorithm enjoy broad uses in both classical and quantum algorithms. In this paper, we propose quantum circuits for GCD computation with $O(n \log n)$ depth with $O(n)$ ancillae. Prior circuit construction needs $O(n^2)$ running time with $O(n)$ ancillae. The proposed construction is based on the binary GCD algorithm and it benefits from log-depth circuits for 1-bit shift, comparison/subtraction, and managing ancillae. The worst-case gate count remains $O(n^2)$, as in traditional circuits.
\end{abstract}

\maketitle

\section{Introduction}
The development and optimization of specific quantum circuits is primarily viewed from the perspective of quantum algorithms in the sense that many quantum models of computation are defined in terms of quantum circuits. In this context, circuit blocks arising in specific quantum algorithms deserve particular attention. Such blocks sometimes implement well-known classical algorithms, but must ensure reversibility, judicious use of ancillae, the restoration of pre-initialized 0 values, and reasonable resource optimization.

Circuit blocks studied in this work encompass GCD computations and variants of the Euclidean Algorithm, which enjoy broad uses in both classical and quantum algorithms.
Classical modular-inverse computations and continued-fraction expansions use similar algorithms. Reversible GCD circuits have been successfully used in quantum algorithms for extracting square-free factors of large integers using Gauss sums \cite{Li2012} and solving Pell's equation \cite{Hallgren:2007}. These algorithms offer significant quantum speed-up. GCD circuits also form the core of algorithms for number-factoring based on Gauss sums \cite{wolk2012}, but these algorithms have been less competitive than other techniques so far \cite{Jones2007,wolk2011}. Other applications include elliptic-curve arithmetic and solutions of the discrete-logarithm problem \cite{Kaye:2005,Proos:2003}. GCD circuits are also attractive as benchmarks for quantum arithmetics, as they are smaller than modular exponentiation circuits \cite{modmultqic}.

In this paper, we propose $O(n \log n)$-depth, $O(n^2)$-size quantum circuits for GCD computation with $O(n)$ ancillae. Prior constructions result in $O(n^2)$ running time with $O(n)$ ancillae \cite{Li2012,Proos:2003}. The remaining part of this paper is organized as follows. We introduce background concepts on quantum circuits in Section \ref{sec:background}. In Section \ref{sec:GCD}, theoretical background for GCD computation is discussed. This section includes an introduction of the simple Euclidean algorithm and its extended version as well as the binary GCD algorithm which is particularly used in this paper. Prior circuit structures are reviewed in Section \ref{sec:prior}. The $O(n\log n)$-depth circuit structure for GCD computation is proposed in Section \ref{sec:approach}, and Section \ref{sec:conclusion} concludes the paper.

\section{Background on Quantum Circuits} \label{sec:background}
A quantum bit (\emph{qubit}) can be treated as a mathematical object that represents a quantum state with two basic states $|0\rangle$ and $|1\rangle$. It can carry a linear combination $|\psi\rangle = \alpha|0\rangle+\beta|1\rangle$ of its basic states, called a \emph{superposition}, where $\alpha$ and $\beta$ are complex numbers and $|\alpha|^2$+$|\beta|^2$=1.

A matrix $U$ is \emph{unitary} if $UU^\dag=I$ where $U^\dag$ is the conjugate transpose of $U$ and $I$ is the identity matrix. An $n$-qubit \emph{quantum gate} performs a $2^n\times2^n$ unitary operation $U$ on $n$ qubits in a specific period of time. For a gate $g$ with a unitary matrix $U_g$, its inverse gate $g^{-1}$ implements the unitary matrix $U_g^{-1}$.
Two gates can be executed in parallel if they share neither control(s) nor target(s).
Given any unitary gate $U$ over $m$ qubits, a controlled-$U$ gate with $k$ control qubits can be defined as an $(m+k)$-qubit gate that applies $U$ on the $m$ qubits if and only if all $k$ control qubits are $\ket{1}$. Additionally,
\begin{itemize}
  \item [$\bullet$] A \emph{multiple-control Toffoli gate} C$^m$NOT $(x_1, \cdots, x_{m+1})$ passes the first $m$ qubits unchanged. These qubits are referred to as \emph{controls}. This gate flips the value of $(m+1)^{th}$ qubit if and only if each positive (negative) control line carries the 1 (0) value. For $m=0,1,2$ the gates are called NOT, CNOT, and Toffoli, respectively.
  \item [$\bullet$] A {\em multiple-control Fredkin gate} Fred$(x_1, x_2, \cdots,x_{m+2})$ has two target lines $x_{m+1},x_{m+2}$ and $m$ control lines $x_1, x_2, \cdots,x_m$. The gate interchanges the values of the targets if the conjunction of all $m$ positive (negative) controls evaluates to 1 (0).
      For $m=0,1$ the gates are called SWAP and Fredkin, respectively.
\end{itemize}

\begin{figure}[tb]
\centering
\scriptsize
\scalebox{.9}{
\input{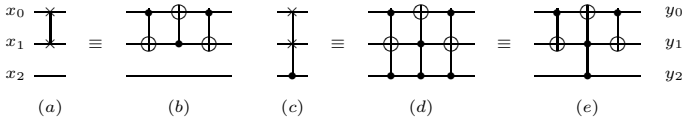}
}
\caption{SWAP gate (a) can be constructed by 3 CNOTs (b). A conditional SWAP (Fredkin) (c) can be implemented by three Toffoli gates (d) or one Toffoli and two CNOTs (e). \label{Fig:swap}}
\end{figure}

In all circuit diagrams, horizontal lines are variables, vertical lines are gates, and time flows left to right. Additionally, $\bullet$ (or $\circ$) is used for conditioning on the qubit being set to value `1' (or `0'), $\oplus$ is used to denote target line of a multiple-control Toffoli gate, and $\times$ is used on qubits of a SWAP (or a controlled SWAP) gate. Fig. \ref{Fig:swap}-a shows a SWAP gate which can be implemented by three CNOT gates as shown in Fig. \ref{Fig:swap}-b. Adding one control to SWAP gate (Fig. \ref{Fig:swap}-c) results in a Fredkin gate which can be implemented with three Toffoli gates (Fig. \ref{Fig:swap}-d) or one Toffoli gate and two CNOTs (Fig. \ref{Fig:swap}-e).

The lines which are added to a quantum circuit are named \emph{ancillae}.\footnote{`ancilla' means `supporting' in Latin.} We use zero-initialized ancillae in this work. The zero-initialized ancillae may be modified inside a given circuit, but should be returned to zero at the end of computation to be reused. The number of qubits, which include both main qubits and ancillae registers, are very limited in current quantum technologies.

\section{Greatest Common Divisor} \label{sec:GCD}
\begin{figure}[tb]
\centering
\scriptsize
\input{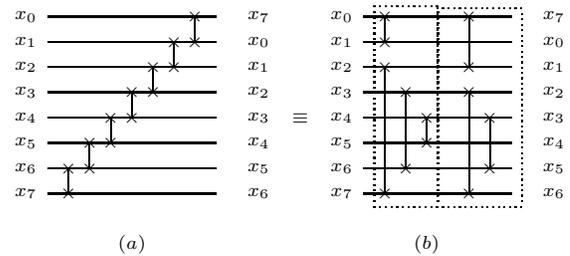}
\caption{ \label{Fig:shift1}  Circular 1-bit shift on 8 qubits. (a) Straightforward implementation. (b) Constant-depth implementation. All SWAP gates in dashed boxes can be executed in parallel.}
\end{figure}

Algorithms discussed in this paper perform integer arithmetic which can be described with C/C++ operators.
\begin{itemize}
\item [$\bullet$] / for integer division, e.g., 10 / 6 = 1
\item [$\bullet$] \% for the remainder operation, e.g., 10 \% 6 = 4
\end{itemize}
In particular, $n/2$ shifts the bits of $n$ to the right by one position, and $n\%2=0$ checks if $n$ is even. As illustrated in Fig. \ref{Fig:shift1}-a, the $n/2$ operator (1-bit shift) can be implemented by a cascade of SWAP gates. This can be verified by exchanging the lines involved in each SWAP gate.

The greatest common divisor (GCD) of two integers $A$ and $B$ can be found by the \emph{Euclidean algorithm} which performs successive division with remainder, given that for $A = Bq+r$ with all positive numbers, $\gcd(A, B) = \gcd(B, r = A\%B)$. 
The {\em extended Euclidean algorithm} additionally finds integers $x$ and $y$ that satisfy B\'ezout's identity $Ax + By = \gcd(A, B)$. For coprime $A$ and $B$, $x$ is the {\em multiplicative inverse} of $A$ modulo $B$, and $y$ is the multiplicative inverse of $B$ modulo $A$. This {\em modular inverse} $A^{-1} \equiv x \pmod{B}$ enjoys applications in various fields including cryptography.

\begin{figure}[tb]
\centering
\scalebox{0.55}{
\input{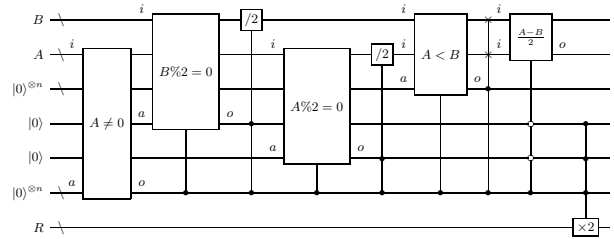}
}
\caption{\label{Fig:B_GCD} An outline of the binary GCD algorithm (one step). A backslash ($\backslash$) on a horizontal line represents a multiqubit bus. The added ancillae are used to evaluate $A<B$, $B\%2=0$, $A\%2=0$, and $A\neq0$, respectively. The last $n$-qubit register is used to hold the value of $R=\gcd(A,B)$ at each step. A $\bullet$ (or $\circ$) on an $n$-qubit register denotes a conditional operation. The input, output, and ancilla registers for each block are specified by `\emph{i}', `\emph{o}', and `\emph{a}'  on the related lines, respectively.
}
\end{figure}

The \textbf{Binary GCD Algorithm} \cite{knuth_art}, also called Stein's algorithm, computes the GCD of two nonnegative integers $a$ and $b$ using subtractions and divisions by two, which are easy to implement in hardware. The algorithm maintains two numbers, starting with $a$ and $b$, but replaces them at every step with a pair that has the same GCD.
The following steps are repeated until either $A = B$ or $A = 0$.
\begin{itemize}
  \item [$\bullet$] If $A\%2=B\%2=0$, $\gcd(A,B)=2 \gcd(A/2, B/2)$%
  \item [$\bullet$] If $A\%2=0=1-B\%2$, $\gcd(A,B) = \gcd(A/2, B)$\\
        If $A\%2=1=1-B\%2$, $\gcd(A,B) = \gcd(A, B/2)$
  \item [$\bullet$] If $A\%2=B\%2=1$,
   then we ensure that $A\geq B$, and use
  $\gcd(A,B)=\gcd \Big(\frac{A - B}{2}, B \Big)$
\end{itemize}
The last branch is performed with a single test $(A < B)$ that controls Fred$(A,B)$, followed by $A=\frac{A - B}{2}$. The binary GCD algorithm is outlined in Fig. \ref{Fig:B_GCD}. In this figure, the register $R$ is used to save the intermediate GCD value at each step. Initially $R=1$ and at each step $R=2 \cdot R$ if $A\%2=B\%2=0$. After the last GCD iteration, $R \cdot B$ computes the result. Note that the comparison blocks may need $n$ zero-initialized ancillae to compute the result, but the resulting value is a single bit.

For $n$-bit numbers, each step takes linear time, given that comparison, subtraction, and circular shift have linear-size circuits. $O(n)$ steps are followed by an $O(n^2)$-gate multiplication $B \cdot R$. Thus, the binary GCD algorithm needs $O(n^2)$ time. On average, it uses 60\% fewer bit operations than the Euclidean algorithm \cite{Akhavi:2000}, but does not improve asymptotic performance. Similar to the extended Euclidean algorithm, an extended binary GCD algorithm is suggested in \cite[p. 338 \& p. 646]{knuth_art} which performs subtractions rather than more general divisions with remainder. The removal of factors of two is irreversible, but can be implemented by circular shifts that move trailing zeros to the most significant bits. Such an arrangement still requires clearing control values. The construction proposed in this work is based on the binary GCD algorithm.

\section{Prior Work} \label{sec:prior}
Our work focuses on GCD and related computations for integers, rather than for polynomials over binary fields \cite{Kaye:2005}. To implement binary GCD by a quantum circuit, \cite{Li2012} used three extra $n$-qubit ancilla registers, see Fig. \ref{Fig:B_GCD} for an outline, to (1) check the termination condition ($A = B$ or $A = 0$) after each step, (2) verify whether $A$ and $B$ are even or not, and (3) check $A < B$. Each step of the algorithm performs even/odd and greater/less comparisons. The maximum possible number of steps should be implemented by explicit circuit blocks, as the actual number of steps depends on $A$ and $B$. This path was pursued in \cite{Li2012} which leads to $O(n^2)$ runtime. In \cite{Proos:2003}, the authors proposed a quantum circuit for the extended Euclidean algorithm with $O(n^2)$ time complexity and $O(n)$ space. Applying the method for the binary extended Euclidean algorithm leads to $7n+\epsilon$ qubits and a running time of $O(n^2)$ \cite{Proos:2003}. The authors did not clear all zero-initialized ancillae which limits the applicability of their techniques.

\section{GCD Circuits with O\lowercase{$(n \log n)$} Depth} \label{sec:approach}
Each step of the binary GCD algorithm includes several data-dependent branches. Given that quantum circuits must work correctly with superposition states, all branches must be implemented explicitly and the longest possible execution trace must be supported. Such a trace includes $n$ steps, each one performs either a single subtraction or divisions by two. In GCD computation, a 1-bit circular right shift can implement the division-by-two operator as the least significant line holds 0 whenever a division-by-two is called. Otherwise, one needs to exclude one line from the rest of computation each time.
When circuit depth is considered, one can use log-depth adder/subtractor circuits with $\Theta(n)$ ancillae \cite{Draper06}, and the conventional implementation of a shift as a sequence of swaps becomes a bottleneck.

\begin{figure}[tb]
\centering
\scriptsize
\scalebox{0.9}{
\input{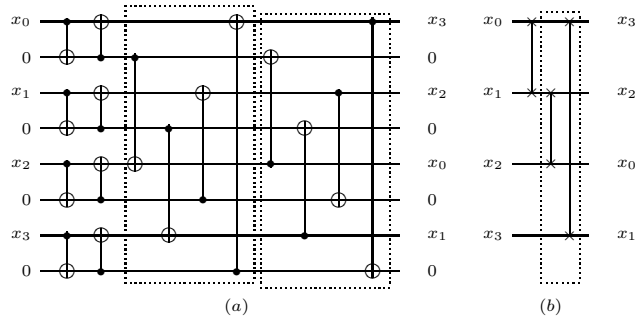}
}
\caption{\label{fig:bit_permutation} Implementation of any fixed bit-permutation by a constant-depth quantum circuit \cite{Moore:2002}. Constructing the permutation cycle $(0,3,1,2)$ by depth 4 with ancillae (a), and with depth 6 without ancillae (b). In (b), each SWAP gate needs 3 CNOT gates for implementation. Gates in dashed boxes can be executed in parallel.}
\end{figure}

To implement a logarithmic-depth circuit for GCD computations, we use ideas from \cite{Moore:2002}, which has not considered GCD computations, but studied parallel quantum circuits. The authors point out that any fixed bit-permutation can be implemented with $O(1)$ depth using $n$ zero-initialized ancilla in four layers --- by copying the bits to ancillae in parallel, canceling the originals, copying the ancillae, and then canceling the ancillae. Consider Fig. \ref{fig:bit_permutation}-a which illustrates the permutation cycle $(0,3,1,2)$ with 4 ancillae. This circuit transforms $x_0$ to $x_3$, $x_3$ to $x_1$, $x_1$ to $x_2$, and finally $x_2$ to $x_0$. On the other hand, the depth of six layers can be achieved with no ancillae by dealing with each cycle individually and decomposing it into a product of two sets of disjoint swaps. Consider a $k$-cycle\footnote{Let $A={1, 2, 3, \cdots, m}$. A \emph{k-cycle} is a permutation $f$ for which there exists an element $x$ in $A$ such that $x, f(x), f^2(x), ..., f^k(x) = x$ are the only elements moved by $f$. In particular, a \emph{transposition} or 2-cycle is a permutation which exchanges two elements and keeps all others fixed.}
$\sigma_k = (1,2,3,4,5,...,k)$. Then for $\pi_k =(1,2) (k,3) (k-1,4) ... $ note that $\pi_k = \pi_k^{-1}$. If $k$ is odd, $\pi_i$ will have one fixed point, but it can anyway be implemented by $k/2$ parallel swaps. Furthermore, note that $\rho_k = \sigma_k \pi_k$ has a similar cycle structure and can also be implemented by parallel swaps. Therefore, by implementing $\rho_k$ and $\pi_k$ with disjoint swaps, we implement $\sigma_k$ in constant depth. Fig. \ref{fig:bit_permutation}-b illustrates the permutation cycle in Fig. \ref{fig:bit_permutation}-a without ancillae. In $(b)$, the first two SWAP gates construct the permutation $(0,1)(1,2)$ and the third SWAP constructs $(0,3)$. Following this path, a depth 2 single-bit circular shift is shown in Fig. \ref{Fig:shift1}-b which includes the transpositions $(0,1)(2,7)(3,6)(4,5)(0,2)(3,7)(4,6)$.

\begin{figure}[t]
\centering
\scriptsize
\scalebox{0.8}{
\input{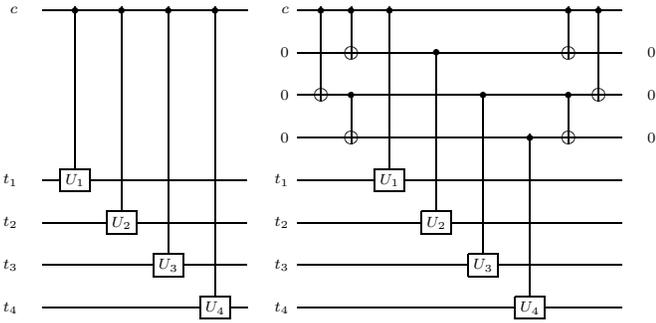}
}
\begin{center}
\caption{\label{fig:shared_bit} $\log n$-depth implementation of shared control (fanout) \cite{Moore:2002}.}
\end{center}
\end{figure}

The work in \cite{Moore:2002} points out that $\sim n$ gates controlled by a shared bit (fanout) cannot be applied in parallel directly, but illustrates a straightforward technique that copies the control value to $n$ ancillae with depth $\log n$ and clears the ancillae after their use. This approach is illustrated in Fig. \ref{fig:shared_bit} where the initial and final CNOT gates are used to prepare and clear the added ancillae, respectively. This adds $2\log n+1$ latency. However, the main circuit block which includes applying conditional unitaries is parallelized to depth 1.

\begin{figure}[tb]
\centering
\scalebox{0.5}{
\input{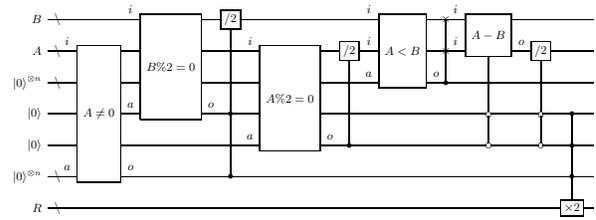}
}
\caption{\label{Fig:B_GCD2} Restructuring the binary GCD algorithm of Fig. \ref{Fig:B_GCD} (one step). The input, ancilla, and output registers for each block are specified by `\emph{i}', `\emph{a}', and `\emph{o}' on the related lines, respectively.
}
\end{figure}

Following Fig. \ref{Fig:B_GCD}, each step of the binary GCD algorithm may include a single conditional subtraction, and/or a single-bit conditional shift. The $A\%2=0$ and $B\%2=0$ blocks can be implemented unconditionally since they either check whether $A$ and/or $B$ are even or not without modifying the values of $A$ and $B$ registers. Similarly, $A<B$ can be computed unconditionally. The conditional 1-bit shift on $A$ when $A\%2=0$ can also be applied even if $A=0$. This simplifies the second circular 1-bit shift operation in Fig. \ref{Fig:B_GCD}. To implement conditional $\frac{A-B}{2}$, note that one of the conditionals is on $A=0$. If $A=0$, then $A\%2=0$, and $\frac{A-B}{2}$ is not applied. Accordingly, $\frac{A-B}{2}$ can be computed with a single conditional. The result of these optimizations is shown in Fig. \ref{Fig:B_GCD2}. Additionally,

\begin{itemize}
  \item [$\bullet$] The unconditional comparison $A < B$ and $A\neq0$ can be implemented by circuits with logarithmic depth with $O(n)$ cleared ancillae \cite{Draper06}.
  \item [$\bullet$] The conditional subtraction $A-B$ can be implemented by a circuit with logarithmic depth with $O(n)$ cleared ancillae. This can be done by following the circuit structure in \cite{Draper06}, and replacing CNOT and NOT gates on output lines by Toffoli and CNOT gates, respectively.
  \item [$\bullet$] Circuit for $A\%2=0$ (and $B\%2=0$) includes a single CNOT conditioned on the last bit of $A$ (and $B$).
  \item [$\bullet$] Swapping two $n$-qubit registers $A$ and $B$ can be done in one step by applying $n$ SWAP gates on disjoint qubits in parallel. Conditional Fred$(A,B)$ can be implemented by $\log n$ depth with $n$ ancillae --- a log-depth circuit to replicate the conditional on $n$ ancillae and a circuit with depth 1 for Fred$(A,B)$. All ancillae can be cleared.
  \item [$\bullet$] Unconditional bit shift can be implemented with a constant-depth circuit. For conditional shift, one can use $n$ ancillae to replicate the control in $O(\log n)$ time. Accordingly, conditional shift can be parallelized to $O(\log n)$ depth. All ancillae can be cleared since the conditional remains unchanged.
\end{itemize}

\begin{table*}[t]
\caption{\label{tab:values} Size, depth, and ancillae in different circuit blocks. The number of CNOT and Toffoli gates are reported separably as a [\#CNOT;\#Toffoli] pair. For comparison and subtraction blocks we used the method in \cite{Draper06}. In those cases, the number of ones in the binary expansion of $n$ is represented by $w(n)$. Prior constructions \cite{Proos:2003,Li2012} use linear-size and linear-depth circuits with $O(n)$ ancillae for each step of the GCD computation where we use linear-size and log-depth with $O(n)$ ancillae.}
\scriptsize
 \begin{tabular}{|l|l|l|}
  \hline
Block & Characteristics & Reference\\
  \hline
   \multirow{3}{*}{Comparison} & \multirow{1}{*}{Size: $[2n -2 ;6n - w(n - 1) - 2  \floor{\log (n - 1)} - 7 ]$} & \multirow{3}{*}{\cite{Draper06}}\\
   & \multirow{1}{*}{Depth: $[2 ;2  \floor{\log n} + 5$]} & \\
   & \multirow{1}{*}{Ancillae: $2n -  \floor{\log (n - 1)} - 3$} & \\
   \hline
  \multirow{3}{*}{Conditional subtraction}     & Size: $[2n;14n - 11]$ & \multirow{3}{*}{\cite{Draper06}}\\
  & Depth: $[2;3\floor{\log(n-1)} + \floor{\log{n-1 \over 3}} + 16]$ & \\
  & Ancillae: $2n-2$  &  \\
  \hline
  \multirow{3}{*}{Conditional 1-bit circular shift}     &Size: $[4n-2;n-1]$  & \multirow{3}{*}{This work}\\
  &  Depth: $[2\ceil{\log n}+4;2]$ & \\
  & Ancillae: $n$ & \\
  \hline
  \multirow{3}{*}{Conditional SWAP}  & Size: $[4n-2;n-1]$ & \multirow{3}{*}{This work}\\
  & Depth: $[2 \ceil{\log n}+4;2]$ & \\
  & Ancillae: $n$ & \\
\hline
\end{tabular}
\end{table*}

Table \ref{tab:values} reports the values of gate count and circuit depth for different circuit blocks. In this table, the numbers of CNOT and Toffoli gates are reported independently as a [\#CNOT; \#Toffoli] pair. Values for comparison and conditional subtraction can be obtained by following the circuit structures, depths, and sizes given in \cite{Draper06} and the notes above. For conditional SWAP, note that $2n$ CNOT gates (with depth $2\ceil{\log n}$) are used to prepare and clear the ancilla register. Each Fredkin gate can be implemented by two CNOT and one Toffoli gates as illustrated in Fig \ref{Fig:swap}-e, and there are $n$ parallel Fredkin gates in total. Therefore, circuit depth can be computed as $2\ceil{\log n}+2$ CNOTs, and one Toffoli. Similarly, circuit size is $4n$ CNOTs, and $n$ Toffoli gates.
To count the number of gates for conditional 1-bit circular shift, note that $2n$ CNOTs (with depth $2\ceil{\log n)}$) are used to prepare and clear ancillae and the remaining $n-1$ Fredkin gates can be implemented with constant depth (i.e., 4 CNOT and 2 Toffoli gates) and linear size (i.e. $2n-2$ CNOT and $n-1$ Toffoli gates). Altogether, the conditional 1-bit circular shift circuit needs $4n-2$ CNOT and $n-1$ Toffoli gates with depth $2\ceil{\log n)}+4$ CNOT and 2 Toffoli Gates. Considering the values given in Table \ref{tab:values} and the circuit structure in Fig. \ref{Fig:B_GCD2} reveals that each step of the GCD computation can be implemented by a log-depth and linear-size circuit.

To compute the final GCD, a multiplication $R \cdot B$ is applied after all steps where $R$ is a power of two. Multiplication by $R$ can be done by a circular shift.\footnote{To implement GCD by a quantum circuit, the method in \cite{Kaye:2005} implements a circular shift by $2^i$ with $i$ blocks of single-bit circular shifts and uses a linear-size circuit for a single-bit shift. However, this is inefficient as any permutation of $n$ qubits can be implemented by a constant-depth circuit \cite{Moore:2002}.} Since $R$ value is computed during GCD iterations, we use $n$ controlled-shifts by $2^i$ ($i<n$).\footnote{For a known $R$ value, a constant-depth circuit suffices.} These power-of-two shifts can be performed in any order, but the conventional quantum-circuit model does not allow parallel execution of gates operating on the same qubits. Since $R$ is a power of two in the GCD computation, only one of the controlled shifts will be applied. Hence, all controlled power-of-two shifts may be applied simultaneously on the same targets. A controlled shift operation can be implemented in $O(\log n)$ depth with $O(n)$ ancillae. Accordingly, the last multiplication of $B$ by $R$ can be implemented with a log-depth, quadratic-size circuit.\footnote{Even if all controlled shift operations needs to be applied in consequence, the final multiplication circuit has $O(n\log n)$ depth with $O(n^2)$ size. Since this multiplication should be applied once, it does not affect the total depth of the the proposed GCD computation circuit --- which is $O(n\log n)$.}

To count ancillae, note that all computational ancillae are cleared inside each block. After the final multiplication block for $R \cdot B$, one can copy (in log depth) the final GCD result to another $n$-qubit zero-initialized register and apply the whole circuit (except for copying the result) in reverse order to recover $A$, $B$, and zero-initialized ancillae. Given that all components use $O(n)$ ancillae (see Table \ref{tab:values}), the total number of ancillae remains linear.

Considering the worst-case number of iterations $n$ to find GCD of two $n$-bit numbers $A$ and $B$, binary GCD computation can be implemented with a $O(n \log n)$-depth, $O(n^2)$-size quantum circuit and $O(n)$ ancillae.

\section{Conclusion} \label{sec:conclusion}
We demonstrated reversible controlled circular-shift circuits with $\log n$ depth and $\Theta(n)$ ancillae. Using these circuits, we proposed $\Theta(n \log n)$-depth quantum circuits for GCD computation.

The Euclidean algorithm finds the greatest common divisor in $O(n^2)$ time. However, it is unknown whether this can be accomplished in $O(\log^{c_1} n)$ time using $O(n^{c_2})$ parallel processors (for constants $c_1,c_2$). Notably, parallel algorithms faster than the Euclidean algorithm have been proposed. The fastest known deterministic classical algorithm solves the problem in $O(n/\log n)$ time with $n^{1+\epsilon}$ processors \cite{ChorG90}. We do not try to make such parallel GCD constructs reversible, and these techniques require significant overhead, including many ancillae and large circuits. Finding a sharper bound on quantum-circuit depth for GCD computation using a reasonable number of gates and ancillae is an interesting open question.

\noindent
{\bf Acknowledgments.}
IM's work was sponsored in part by the  Air Force Research Laboratory
under agreement FA8750-11-2-0043.

\end{document}